\documentclass[lettersize,journal]{IEEEtran}
\usepackage{amsmath,amsfonts}
\usepackage{algorithmic}
\usepackage{algorithm}
\usepackage{array}
\usepackage[caption=false,font=normalsize,labelfont=sf,textfont=sf]{subfig}
\usepackage{textcomp}
\usepackage{stfloats}
\usepackage{url}
\usepackage{verbatim}
\usepackage{graphicx}
\usepackage{cite}
\usepackage{enumitem}
\usepackage{hyperref}
\usepackage{makecell}
\usepackage{etoolbox}
\robustify\bfseries
\usepackage{amssymb}
\usepackage{tabularx}
\usepackage{wasysym}
\usepackage{color,soul}
\usepackage{balance}
\usepackage{booktabs}
\usepackage{multirow}
\usepackage{siunitx}
\usepackage{arydshln}

\def\RR{{\mathbb R}}
\usepackage{caption}
\usepackage{amssymb}%
\usepackage{pifont}%
\newcommand{\xmark}{\ding{55}}%

\usepackage{tikz}

\usepackage{flushend}

\newcommand{\ZQHL}[1]{{#1}}

\newcommand{\PJHL}[1]{{#1}}

\begin{document}

\title{Listen to Extract:\\Onset-Prompted Target Speaker Extraction}

\author{Pengjie Shen, Kangrui Chen, Shulin He, Pengru Chen, Shuqi Yuan, He Kong, Xueliang Zhang,\\and Zhong-Qiu Wang
\thanks{Manuscript received on May 8, 2025; \PJHL{revised Oct. 6, 2025}; accepted Nov. 3, 2025. This work was done while P. Shen was visiting student
at Southern University of Science and Technology. \textit{(Corresponding author: Zhong-Qiu Wang).}
}
\thanks{P. Shen and X. Zhang are with the Department of Computer Science, Inner Mongolia University, Hohhot 010021, China (e-mail: shenpengjie@mail.imu.edu.cn; cszxl@imu.edu.cn).}
\thanks{
H. Kong is with the School of Automation and Intelligent Manufacturing,
Southern University of Science and Technology, Shenzhen 518055, China (e-mail: kongh@sustech.edu.cn).}
\thanks{
K. Chen, S. He, P. Chen, S. Yuan and Z.-Q. Wang are with the Department of Computer Science and Engineering, Southern University of Science and Technology, Shenzhen 518055, China (e-mail: chenkr2021@mail.sustech.edu.cn;
\{hesl,wangzq3\}@sustech.edu.cn;
wang.zhongqiu41@gmail.com).}
}

\markboth{}
{Shell \MakeLowercase{\textit{et al.}}: A Sample Article Using IEEEtran.cls for IEEE Journals}

\maketitle

\begin{abstract}
\ZQHL{We propose \textit{listen to extract} (LExt), a highly-effective while extremely-simple algorithm for monaural target speaker extraction (TSE).}
Given an enrollment utterance of a target speaker, LExt aims at extracting the target speaker from the speaker's mixed speech with other speakers.
For each mixture, LExt concatenates an enrollment utterance of the target speaker to the mixture signal at the waveform level, and trains deep neural networks (DNN) to extract the target speech based on the concatenated mixture signal.
The rationale is that, this way, an artificial speech onset is created for the target speaker and it could prompt the DNN (a) which speaker is the target to extract; and (b) spectral-temporal patterns of the target speaker that could help extraction.
This simple approach produces strong TSE performance on multiple public TSE datasets including WSJ0-2mix, WHAM! and WHAMR!.
\end{abstract}


\section{Introduction}

\IEEEPARstart{I}{n} many artificial intelligence and machine learning applications, the sensors inevitably record a mixture of target and non-target signals.
The recorded non-target signals often pose tremendous difficulties in the perception and understanding of the target signals.
In this case, how to extract and enhance the target signal(s) is an important research problem to study.
One example, in audio signal processing, is speaker separation (\textit{a.k.a.}, the cocktail party problem) \cite{E.Cherry1953}, \cite{McDermott2009}, \cite{WDLreview}, where, in a noisy-reverberant room with multiple speakers talking concurrently, the task is to extract each of the target speakers of interest based on the mixtures recorded in the room.
In the past decade, dramatic progress has been made in speaker separation \cite{WDLreview}, \cite{Kolbak2017}, \cite{Hershey2016}, thanks to the rapid development of deep learning.
Some studies train deep neural networks (DNN) to separate all the speakers in the mixture via permutation invariant training (PIT) \cite{Hershey2016, Kolbak2017}, where the input to the DNN is a multi-speaker mixture and the DNN is trained, in a supervised way, to estimate all the speakers in the mixture.

Differently, some other studies only aim at target speaker extraction (TSE), which targets at extracting a particular speaker of interest by providing to the DNN some auxiliary cues that can indicate which speaker is the targeted one to extract \cite{Delcroix2020TDSpeakerBeam, Delcroix2023SoundBeam, Elminshawi2022NewInsightsTSE, Ge2022LSpEx, Michelsanti2021, Ochiai2025TSIE, Wang2018TSE, Wang2018kVoiceFilter, Wang2024WeSep, Wang2023MISP, Wu2024MISP, Zmolikova2023SPM, Zmolikova2017, Zmolikova2019SpeakerBeam}.
TSE is an important problem to study, since in many application scenarios one may only care about a particular target speaker of interest.
One approach to realize TSE is to first separate all the speakers, and then identify the speaker estimate corresponding to the target speaker \cite{Ochiai2025TSIE}.
However, as is discussed in \cite{Ochiai2025TSIE}, this approach often produces suboptimal performance compared to TSE for the following reasons:
(a) separating all the speakers is a difficult task, especially when the number of mixed speakers is large;
(b) the separated speakers suffer from permutation ambiguity (i.e., an arbitrary assignment of speaker estimates to true target speakers) and subsequent effort is needed to identify the speaker estimate corresponding to the target speaker;
and (c) the number of speakers is typically unknown and can vary a lot in different mixtures, creating difficulties for separating all the speakers, as speaker counting may not be perfect.

Many auxiliary cues can be utilized for TSE, such as an enrollment utterance or a speaker embedding \cite{Wang2018kVoiceFilter, Zmolikova2019SpeakerBeam}, visual lip movements \cite{Ephrat2018, Gabbay2018}, bone-conduction signals \cite{Wang2022BoneConduction}, speaker-activity timestamps \cite{Delcroix2021}, direction information \cite{Gu2020} and brain signals \cite{Ceolini2020, Pan2024NeuroHeed} of the target speaker.
Among various auxiliary cues, enrollment utterances are typically easy to obtain, as users can easily register their voice to the TSE system by, for example, speaking something to their cell phones.
With this observation, in this paper we investigate TSE given an enrollment utterance of the target speaker.
Previous algorithms in this direction \cite{Michelsanti2021, Ochiai2025TSIE, Wang2024WeSep, Zmolikova2023SPM} usually extract a fixed-length speaker embedding from the enrollment utterance, and use it to condition a speaker extraction network at one or multiple layers to extract the target speaker.
In the past decade, many conditioning mechanisms have been proposed, such as performing concatenation \cite{Wang2018TSE}, summation \cite{Afouras2019, Wang2024WeSep}, multiplication \cite{Delcroix2020TDSpeakerBeam}, cross-attention \cite{Ochiai2019, Sato2021}, and feature-wise linear modulation (FiLM) \cite{Cornell2022Clarity,Hao2024XTFGridNet} operations between the fixed-length speaker embedding and the internal tensors produced by the speaker extraction network.
A popular way to extract the fixed-length speaker embedding is by using pre-trained speaker recognition models \cite{Wang2018kVoiceFilter} such as the x-vector \cite{Snyder2018xvector} and ECAPA-TDNN \cite{Desplanques2020ECAPATDNN} models.
Another way is to train the speaker extraction network with a speaker embedding network through multi-task learning, where the speaker embedding network can be initialized by using a pre-trained speaker recognition model and a speaker classification loss can be attached to maintain the speaker discrimination capabilities of the pre-trained speaker recognition model \cite{Delcroix2020TDSpeakerBeam, Zmolikova2019SpeakerBeam}.
Although these two approaches have been very popular in recent years in TSE research, they have two issues \cite{Michelsanti2021, Ochiai2025TSIE, Zmolikova2023SPM}.
First, the fixed-length speaker embedding may
mismatch the tensors produced inside the speaker extraction network, leading to difficulties in achieving highly-efficient speaker conditioning.
Second, the fixed-length speaker embedding is designed to be capable of discriminating different speakers for speaker recognition, but it may not be optimal for the final TSE task.

To deal with the above issues, a stream of research attempts to not rely on fixed-length speaker embeddings, and instead designs a way to leverage variable-length speaker embeddings so that the temporal patterns and local dynamics of the target speaker in the enrollment utterance can be exploited for TSE.
Xiao \textit{et al.} \cite{Xiao2019} first embed the mixture and the enrollment utterance, obtaining an embedding sequence for the mixture and another for the enrollment utterance, and then a cross-attention operation, which uses each embedding in the mixture embedding sequence as query and the embeddings in the enrollment sequence as key and value, is applied to obtain a new embedding sequence, which, by design, has the same length as the mixture embedding sequence.
Next, the new embedding sequence is combined with the mixture embedding sequence for subsequent DNN modules to extract the target speaker.
Notice that the new embedding sequence can be considered as variable-length speaker embeddings and could be more matched with the mixture embeddings for speaker conditioning, compared with fixed-length speaker embeddings.

This research has motivated subsequent studies such as CIENet \cite{CIENet-c2f,CIENet} and USEF-TSE \cite{Zeng2024USEF-TSE}.
CIENet \cite{CIENet-c2f, CIENet} uses raw short-time Fourier transform (STFT) coefficients as the query, key and value tensors for the cross-attention operation proposed in \cite{Xiao2019}, and in addition employs a stronger speaker extraction network.
USEF-TSE \cite{Zeng2024USEF-TSE}, an improved version of SEF-Net \cite{Zeng2023SEF-Net}, improves Xiao \textit{et al.}'s approach by employing an advanced cross-attention mechanism proposed in TF-GridNet \cite{Wang2023TFGridNet} to obtain variable-length speaker embeddings for speaker conditioning and by using an advanced DNN architecture (i.e., TF-GridNet) for speaker extraction.
USEF-TSE reports strong performance on multiple TSE benchmarks.\footnote{USEF-TSE \cite{Zeng2024USEF-TSE} and its preliminary version SEF-Net \cite{Zeng2023SEF-Net} argue that they are \textit{speaker-embedding-free}, since they do not rely on conventional speaker embeddings, which are usually fixed-length. However, one may consider that a variable-length speaker embedding is extracted in their system. As a result, this paper avoids using the term \textit{speaker-embedding-free}.}

In this context, we propose another TSE algorithm free of fixed-length speaker embeddings named \textit{listen to extract} (LExt).
Compared with existing studies, it is much simpler while obtaining better TSE performance.
It concatenates an enrollment utterance of the target speaker to the input mixture (e.g., by prepending the enrollment utterance to the mixture at the waveform level), and trains a DNN to extract the target speaker based on the concatenated mixture.
The idea is to have the DNN first listen to an example utterance of the target speaker, which may, later on, prompt the DNN to extract another utterance of the target speaker from the observed mixture.
We highlight that LExt does not require customized DNN architectures to realize such a speaker conditioning.
Many existing DNN architectures in speech separation (e.g., TF-GridNet \cite{Wang2023TFGridNet}, SepFormer \cite{Subakan2021} and TF-LocoFormer \cite{Saijo2024LocoFormer}) could be straightforwardly employed, if the DNN (a) has a sufficiently-large receptive field (e.g., realized by self-attention) to see the concatenated enrollment utterance while extracting the target speaker from the input mixture; or (b) has a memory mechanism such as long short-term memory (LSTM) that can memorize what the DNN has listened to in the beginning of the prepended mixture.
In our experiments on multiple public datasets designed for TSE (including WSJ0-2mix \cite{Hershey2016}, WHAM! \cite{Wichern2019} and WHAMR! \cite{Maciejewski2020}), LExt achieves strong TSE performance, despite being very simple. A sound demo is provided in the link below\footnote{See \url{https://zqwang7.github.io/demos/LExt_demo/index.html}.}.

Why would such a simple approach work?
There could be two reasons:
\begin{itemize}[leftmargin=*,noitemsep,topsep=0pt]
\item The first reason is related to \textit{onset-based speaker separation} \cite{Taherian2024Onset}, which aims at separating all the speakers (but is not designed for TSE).
Instead of using PIT to address the permutation ambiguity problem, it orders all the speakers according to their speech onset (e.g., from the earliest to latest) and trains DNNs to predict all the speakers according to this order.
In \cite{Taherian2024Onset}, the authors observe strong performance of this onset-based approach for two-speaker separation.
In an earlier study conducted by the same research group, the authors propose attentive training \cite{Pandey2023AttentiveTraining}, where a DNN is trained to estimate the speaker signal with the earliest speech onset, and the training has been shown very successful.
In LExt, the prepended enrollment utterance manually creates, for the target speaker, a speech onset earlier than all the other speakers.
This earliest onset could make the extraction of the target speaker possible.
\item The second reason could be that, in LExt, at every DNN layer, the hidden representations extracted from the enrollment utterance are \textit{homogeneous} with those extracted from the mixture, simply because they are extracted by exactly the same DNN modules.
This way, the speaker conditioning in LExt could be more effective, compared with just conditioning some layers of the speaker extraction DNN by speaker embeddings, which are extracted by other DNN modules (e.g., a DNN trained for speaker recognition) and hence may be much less homogeneous with the hidden representations produced by the speaker extraction DNN.
In addition, in LExt, implicit speaker conditioning exists at \textit{every layer} of the DNN, as we use exactly the same DNN to process every frame of the concatenated mixture.
This could be a much more fine-grained speaker conditioning mechanism than speaker-embedding-based methods, which usually only perform speaker-embedding-based conditioning at a shallow layer \cite{Michelsanti2021, Ochiai2025TSIE, Zmolikova2023SPM} or at a selected subset of intermediate layers \cite{cornell2023multi, Hao2024XTFGridNet, He2025, Zeng2023SEF-Net}.
\end{itemize}

The remainder of this paper is organized as follows.
Section \ref{related_work} discusses related works, \ref{LExt_description} proposes LExt, \ref{exp_setup} presents experimental setup, \ref{eval_results} reports evaluation results, \ref{limitations} discusses limitations, and \ref{conclusion} draws conclusions.

\section{Related Works}\label{related_work}

LExt is related to existing studies mainly in the following aspects.

\subsection{TSE Methods Not Using Fixed-Length Speaker Embeddings}

The approach proposed by Xiao \textit{et al.} \cite{Xiao2019}, SEF-Net \cite{Zeng2023SEF-Net}, CIENet \cite{CIENet-c2f, CIENet}, and SEF-Net's improved version USEF-TSE \cite{Zeng2024USEF-TSE} are representative TSE algorithms that do not rely on fixed-length speaker embeddings.
We have
described them in the introduction section.
Compared with them, LExt is a much simpler algorithm while obtaining better TSE performance.
This is possibly due to the implicit speaker conditioning mechanism in LExt, which can happen at every layer of the DNN.
In comparison, in, e.g., USEF-TSE \cite{Zeng2024USEF-TSE}, the cross-attention layer, which is the key for speaker conditioning, is placed at a shallow layer, operating on the embeddings obtained by applying a linear encoder to the spectrograms of the mixture and enrollment utterance.
Such a mechanism may not be very effective at speaker conditioning.

\subsection{Leveraging Estimated Onset for TSE}

In \cite{Hao2021WASE}, the speech onset of the target speaker is estimated based on the input mixture and enrollment utterance, and the estimated onset is encoded as a binary vector (which has the same length as the mixture) to improve TSE systems that are based on fixed-length speaker embeddings.
Very differently, LExt creates an artificial onset to realize TSE.

\subsection{Onset-Prompted Overlapped Speech Processing}

There are existing algorithms based on onset-prompting for multi-speaker overlapped speech processing \cite{Kanda2020SOT, Pandey2023AttentiveTraining, Taherian2024Onset}.
One example is in speaker separation, where the task is to separate the mixture of multiple speakers to individual speakers and the DNN model needs to resolve permutation ambiguity (i.e., figuring out how to align estimated speakers with ground-truth target speakers to realize successful training).
Differently from using a popular solution named PIT \cite{Hershey2016, Kolbak2017}, which first aligns estimates with labels before loss computation, the authors in \cite{Pandey2023AttentiveTraining, Taherian2024Onset} propose to order DNN estimates according to the onset information of different speakers.
Strong speaker separation performance is observed.
Another example \cite{Kanda2020SOT} is in multi-speaker automatic speech recognition, where the task is to recognize the speech of each speaker in multi-speaker mixtures.
The authors propose a solution named serialized output training \cite{Kanda2020SOT}, which orders DNN-estimated token sequences according to the onset cues of the mixed speakers.
The rationale of the above studies is that different speakers usually start talking at different time, and such differences in onset can be leveraged by carefully-designed DNN modules to resolve permutation ambiguity.

Different from existing onset-based overlapped speech processing methods \cite{Kanda2020SOT, Pandey2023AttentiveTraining, Taherian2024Onset}, where the goal is to separate or recognize all the mixed speakers, LExt deals with TSE, a task where there is a single target speaker of interest to extract.
In addition, LExt proposes to introduce an \textit{artificial} onset for the target speaker by, e.g., prepending an enrollment utterance to the mixture.
These differences set LExt apart from existing onset-prompted algorithms in overlapped speech processing.

\section{LExt}\label{LExt_description}

Given a noisy-reverberant environment with $C$ speakers, the physical model of the mixture signal captured by a far-field microphone can be formulated, in the time domain, as
\begin{align}\label{physical_model}
y = s + v \in \RR^N,
\end{align}
where $N$ denotes the number of time-domain samples (i.e., signal length), and $y$, $s$, and $v$ respectively represent the captured mixture, target speech by a desired target speaker we aim to extract, and non-target signals.
We use $v$ to absorb all the signals that are not considered as target speech, including the reverberation of the target speaker, environmental noises, and signals produced by the other speakers.
In the task of TSE, an $E$-sample-long enrollment utterance $e\in\RR^E$ is assumed available.
It is an utterance different from the target speech $s$, but is uttered by the same speaker as $s$.
This way, it can indicate who the target speaker is and hence can be utilized to help extract the target speaker from the mixture.
The rest of this section describes our proposed LExt algorithm and its design choices.

\begin{figure}
  \centering  
  \includegraphics[width=8cm]{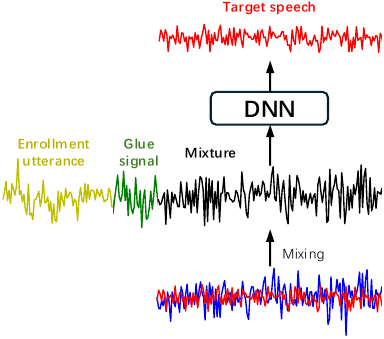}
  \caption{
  Illustration of LExt for TSE. In this example, an enrollment utterance and a glue signal are prepended to the mixture for the DNN to extract the target speaker.
  Best viewed in color.
  \PJHL{
  The mixture signal (shown in black) consists of the target speech (in red) and non-target signals (in blue).
  }
  }
  \label{system_overview}
\end{figure}

\subsection{Listen to Extract (LExt)}

The proposed LExt approach is illustrated in Fig. \ref{system_overview}.
This embodiment of LExt prepends an enrollment utterance $e$ to the mixture $y$.
They are concatenated by using a $G$-sample-long glue signal $g\in\RR^G$.
The augmented mixture signal is fed to a DNN, which is trained to extract the target speaker indicated by the enrollment utterance.

To train the DNN, for each clean and mixture signal pair $<s, y>$\footnote{\PJHL{The notation, $<\cdot,\cdot>$, denotes a signal pair consisting of a mixture input signal and its corresponding target signal.}} used for training, we first augment each of the two signals to create a new pair $<\widetilde{s}, \widetilde{y}>$, where $\widetilde{y}=\big[e; g; y\big]\in\RR^{E+G+N}$ prepends an enrollment utterance $e$ and a glue signal $g$ to the mixture $y$, and $\widetilde{s}=\big[e; g; s\big]\in\RR^{E+G+N}$ prepends the same enrollment utterance $e$ and the same glue signal $g$ to the target speech $s$.
The new pair is then utilized to train the DNN via supervised learning.

\subsection{Length and Values of Glue Signal}

In LExt, a glue signal $g$, described in the previous subsection, is utilized to prompt the DNN the time ranges of the enrollment utterance and the mixture.
Since a longer glue signal increases the length of the augmented mixture and would require more computation, we set it to only $32$ ms long.
On the other hand, for simplicity, we set it to an all-zero signal.

\subsection{Loss Functions}\label{loss_range}

The loss function is defined on the estimated target speech after discarding the predictions in the time range of the concatenated enrollment and glue signals (i.e., on $\hat{s}=\hat{\widetilde{s}}[E+G:]$, where $[E+G:]$ means removing the first $E+G$ samples).
Any off-the-shelves loss functions in TSE can be used.
Our study just chooses the scale-invariant signal-to-distortion ratio (SI-SDR) loss \cite{LeRoux2019}, which is widely-adopted in TSE research.

\subsection{DNN Architectures}

LExt can work with many modern DNN architectures in speech separation and enhancement, as long as the DNN can have a mechanism (such as self-attention) that can see the prepended enrollment utterance (at the very beginning of the augmented mixture), while extracting the target speaker.

In our experiments, we investigate two state-of-the-art separation models for LExt, TF-GridNet \cite{Wang2023TFGridNet} and TF-LocoFormer \cite{Saijo2024LocoFormer}.
Their configurations are described later in Section \ref{dnn_architecture}.
Note that, to enable the speaker-conditioning mechanism in LExt, we do not need to modify those DNN architectures, and instead we just need to change the input and output of those DNN architectures and the DNN could learn to leverage the artificial onset to extract the target speaker.
This property distinguishes LExt from existing TSE methods \cite{Michelsanti2021, Ochiai2025TSIE, Zmolikova2023SPM}, which need to modify the DNN architectures internally and add some internal DNN modules to enable speaker conditioning.

Both networks are trained via complex spectral mapping \cite{Tan2020}, \cite{Wang2020chime}, \cite{Wang2020css} to predict the real and imaginary (RI) components of the target speech based on the RI components of the input signal.

\subsection{Gain Normalization of Enrollment Utterance and Mixture}\label{gain_norm}

We find it beneficial to balance the gain of the mixture $y$ and the enrollment utterance $e$ before concatenating them.

During training, we normalize the sample variance of each signal to $1.0$ before concatenation.
In detail, given the mixture signal $y$, target speech $s$ and enrollment utterance $e$, we normalize each of them in the following way, before creating an augmented pair $<\widetilde{y}, \widetilde{s}>$ for training:
\begin{align}
s &:= s / \sigma(y), \\
y &:= y / \sigma(y), \\
e &:= e / \sigma(e),
\end{align}
where $\sigma(\cdot)$ computes the standard deviation of the signal at the sample level.

At run time, we do the same normalization to $y$ and $e$ before concatenating them for inference.
After obtaining target estimate $\hat{\widetilde{s}}$, we multiply the samples in the time range of the enrollment utterance with $\sigma(e)$, and multiply the samples in the mixture range with $\sigma(y)$ to reverse the normalization.

\subsection{Only Prepend, or Not Only Prepend But Also Append?}

So far, we assume that the enrollment utterance is prepended to the mixture signal.
Alternatively, we can prepend half of the enrollment utterance to the mixture while append the rest.
This could be helpful for DNN architectures such as TF-GridNet \cite{Wang2023TFGridNet} but not for architectures such as TF-LocoFormer \cite{Saijo2024LocoFormer}, since TF-GridNet relies heavily on bi-directional LSTM (BLSTM) modules, which may have a limited receptive field \cite{Courville2016}, while TF-LocoFormer has many self-attention blocks, which lead to a large receptive field.

\subsection{Removing Silences in Enrollment Utterance}

For each enrollment utterance, we use a pre-trained speech activity detector\footnote{\url{https://kaldi-asr.org/models/m4}}, which is based on the Kaldi toolkit, to remove the silences in the enrollment utterance, as they are not informative about the speaker characteristics of the target speaker.
In addition, removing them can reduce the computation of LExt, as the length of the enrollment utterance is reduced.
For the identified silence segments, we simply discard them and splice together non-silent segments.

\section{Experimental Setup}\label{exp_setup}

We validate the effectiveness of LExt at TSE based on public datasets, including WSJ0-2mix, WHAM! and WHAMR!, all of which have been widely-adopted in previous TSE studies.
This section describes the datasets, system configurations of LExt, evaluation metrics, and baseline systems.

\subsection{Datasets}

The WSJ0-2mix \cite{Hershey2016}, WHAM! \cite{Wichern2019} and WHAMR! \cite{Maciejewski2020} are all originally designed for talker-independent multi-speaker separation.
They are later modified for TSE, by specifying each one of the speakers in each mixture in turn as the target speaker and specifying an enrollment utterance for the target speaker.
For all the three datasets, we use the same enrollment utterances\footnote{See \url{https://github.com/ZBang/USEF-TSE/tree/main/data/test/}.} as earlier studies for evaluation.
The average length of the enrollment utterances is $\sim$$7.3$ seconds.
For all the datasets, we use the \textit{min} and $8$ kHz version.

\textbf{WSJ0-2mix} \cite{Hershey2016} is so far the most popular dataset to evaluate monaural talker-independent speaker separation algorithms in anechoic conditions.
It consists of $20,000$ ($\sim$$30.4$ h), $5,000$ ($\sim$$7.7$ h) and $3,000$ ($\sim$$4.8$ h) two-speaker mixtures in its training, validation and test sets, respectively.
The clean source signals are sampled from the WSJ$0$ corpus.
The speakers for training and validation do not overlap with the speakers for testing.
The two utterances in each mixture are fully-overlapped, and their relative energy level is uniformly sampled from the range $[-5, 5]$ dB.

\textbf{WHAM!} \cite{Wichern2019}, buiding on the WSJ0-2mix dataset, introduces environmental noises, extending the two-speaker separation task in WSJ0-2mix to noisy conditions.
The noise dataset used for mixing consists of $80$ hours of real-recorded signals in urban areas such as coffee shops, restaurants, bars, parks and office buildings.
For each two-speaker mixture in WSJ0-2mix, a randomly-selected noise signal is added such that the signal-to-noise ratio (SNR) of the lounder speaker and the noise is equal to a value randomly sampled from the range $[-6, 3]$ dB.
WHAM! consists of $20,000$ ($\sim$$30.4$ h), $5,000$ ($\sim$$7.7$ h) and $3,000$ ($\sim$$4.8$ h) noisy two-speaker mixtures in its training, validation and test sets, respectively.

\textbf{WHAMR!} \cite{Maciejewski2020} is used to validate our algorithms in noisy-reverberant conditions.
It is based on the two-speaker mixtures in WSJ0-2mix \cite{Hershey2016} but reverberates each clean speaker source and adds non-stationary noises.
In each mixture, the reverberation time (T60) is randomly sampled from the range $[0.2, 1.0]$ s, the SNR between the louder speaker and noise is drawn from $[-6, 3]$ dB, relative energy level between the two speakers from $[-5, 5]$ dB, and speaker-to-array distance from $[0.66, 2.0]$ m.
There are $20,000$ ($\sim$$30.4$ h), $5,000$ ($\sim$$7.7$ h) and $3,000$ ($\sim$$4.8$ h) binaural mixtures for training, validation and testing, respectively.
The direct-path signal of the target speaker at the first microphone is considered as the target signal for training and as the reference signal for metric computation.
We only use the first microphone for training and evaluation.
In this TSE task, the model needs to jointly reduce room reverberation, environmental noises and competing speech.

\subsection{DNN Configurations}\label{dnn_architecture}

We consider two DNN architectures for LExt, including TF-GridNet \cite{Wang2023TFGridNet} and TF-LocoFormer \cite{Saijo2024LocoFormer}, both of which operate in the time-frequency (T-F) domain and are representative speech separation models reporting strong separation performance in recent supervised speech separation benchmarks.
TF-LocoFormer is a dual-path DNN architecture like TF-GridNet but replaces BLSTMs with transformer blocks.
The transformer blocks can easily see the concatenated enrollment utterance via its self-attention mechanism.

For each architecture, we investigate two configurations, denoted as \textit{V1} and \textit{V2}.
The \textit{V1} version uses less computation for faster experimentation.
Following the symbols defined in Table I of the TF-GridNet paper \cite{Wang2023TFGridNet}, for \textbf{TFGridNetV1} we set the hyper-parameters to $D=128, B=4, I=1, J=1, H=200, L=4$ and $E=16$, and, for \textbf{TFGridNetV2}, we set them to $D=128, B=6, I=1, J=1, H=256, L=4$ and $E=16$.
Following the TF-LocoFormer paper \cite{Saijo2024LocoFormer}, we employ its small version, denoted as \textbf{TFLocoFormerV1}, and its medium version, denoted as \textbf{TFLocoFormerV2}.
Please do not confuse the symbols used by TF-GridNet and TF-LocoFormer with the ones defined in this paper.

\subsection{Miscellaneous Configurations of LExt}

The model is optimized by using the Adam optimizer for TF-GridNet, while by using the AdamW optimizer for TF-LocoFormer, following the suggestions by the original authors.
For STFT, the square root of Hann window is used as the analysis window, and for TF-GridNet, the window size and hop size are respectively set to $16$ and $8$ ms, while for TF-LocoFormer, they are respectively set to $32$ and $16$ ms to speed up training.

For each training mixture, at each epoch, we randomly choose one of the mixed speakers as the target speaker, and select another utterance of the target speaker as the enrollment utterance.
We train LExt by using fixed-length signal segments sampled from the original mixtures and enrollment utterances.
The length of the sampled mixture segments is $4.0$ seconds, while the length of the sampled enrollment segments is a tunable hyper-parameter.
During training, the length of the enrollment segments and that of the mixture segments are both fixed.
At inference time, the length of the enrollment utterance is fixed and configured the same as the length used during training, while the mixture is in its full length.
\PJHL{
Our motivations for designing the enrollment as fixed-length are two-fold.
First, since LExt concatenates the enrollment directly with the mixture as the network input, using the original, full-length utterance as the enrollment would make the overall input sequence to the network excessively long. This not only increases computation but also risks the model forgetting or under-utilizing the enrollment information, especially in scenarios where the original enrollment utterance itself is long.
Second, we found that adopting a fixed-length enrollment improves training stability by reducing the variability introduced by using enrollment segments of different lengths. This ensures more consistent training conditions and mitigates uncertainties caused by highly-variable input lengths.
}

\subsection{Evaluation Metrics}

Our evaluation metrics follow existing TSE studies.
We use SI-SDR \cite{LeRoux2019} as the main evaluation metric.
We additionally employ BSS-Eval SDR \cite{Vincent2006a} and narrow-band percetual evaluation of speech quality (PESQ) \cite{Rix2001}.
All of them are widely-adopted in TSE research.
For SI-SDR and SDR, we report scores in SI-SDR improvement (SI-SDRi) and SDR improvement (SDRi) over unprocessed mixtures.
The scores are averaged over the mixtures in the test set.

\subsection{Baseline Systems}\label{baseline}

There are two main categories of baseline systems, one based on variable-length speaker embeddings and another based on fixed-length speaker embeddings.

We compare LExt with USEF-TSE \cite{Zeng2024USEF-TSE}, which leverages variable-length speaker embeddings for speaker conditioning and has been described in the introduction section.
It builds upon TF-GridNet \cite{Wang2023TFGridNet} for TSE.
We emphasize that TFGridNetV2 used in this paper is the same as the TF-GridNet architecture used in the original USEF-TSE paper, except for the following two differences.
First, USEF-TSE has an additional speaker conditioning module, based on cross-attention, to enable TF-GridNet to perform TSE.
Second, in USEF-TSE, the embedding dimension of each T-F unit is set to $256$ (as it stacks the embedding of the mixture and the variable-length speaker embedding, both of which are $128$-dimensional at each T-F unit), while, in LExt, it is $128$.

We also compare LExt with TSE systems based on fixed-length speaker embeddings.
We employ such a system suggested by the WeSep toolkit \cite{Wang2024WeSep}, where a speaker embedding network (i.e., ECAPA-TDNN \cite{Desplanques2020ECAPATDNN}) is used to extract a fixed-length speaker embedding from the enrollment utterance and the speaker embedding is used to condition a TF-GridNet based speaker extraction network at multiple layers.
Following \cite{Wang2024WeSep}, we remove the speaker classification loss of ECAPA-TDNN, and the speaker embedding network and the speaker extraction network are jointly trained to optimize the SI-SDR loss.
See Section 3 of the WeSep paper \cite{Wang2024WeSep} for more details.

On the other hand, since LExt is evaluated on public datasets, its evaluation results can be directly compared with many existing ones to show its effectiveness.

\section{Evaluation Results}\label{eval_results}

We first present ablation results based on the WSJ0-2mix dataset to validate various design choices of LExt, and then report the performance of LExt on multiple public benchmarks.

\subsection{Effects of Length of Enrollment Utterance}

In Table \ref{results_enrollment_length}, we compare the results of using enrollment utterances with various durations for speaker conditioning.
Each result is obtained by fixing the lengths of all the enrollment utterances to $E$ seconds during training and testing, where $E\in\{0.1, 0.25, 0.5, 1.0, 2.0, 4.0, 6.0\}$ seconds.

After the silence segments are removed using speech activity detection (SAD), some enrollment utterances are shorter than $E$ seconds.
In this case, we pad zeros to the left, rather than to the right, of the enrollment utterance before prepending it to the mixture, considering that, this way, the active speech in the enrollment utterance can be closer to the mixture and this padding strategy could benefit models such as TF-GridNet which relies heavily on BLSTM-based sequential modules.
After removing the silence segments using SAD, for the enrollment utterances still longer than $E$ seconds, we randomly sample an $E$-second segment in each training epoch for model training, while we always use the first $E$-second segment for evaluation.

From the results in Table \ref{results_enrollment_length}, we observe that LExt does not require very long enrollment utterance.
For example, using TFGridNetV1 and only $0.5$-second-long enrollment speech can yield $22.0$ dB SI-SDRi. This result is already better than many existing TSE systems (shown later in Table \ref{results_compare_with_others}), which leverage enrollment utterances in their full length (on average $\sim$$7.3$ seconds) for speaker conditioning.
For TFGridNetV1, the SI-SDRi result becomes better when the length of the enrollment utterance increases from $0.1$ to $4.0$ seconds.
This is reasonable as longer enrollment utterance can offer more cues about the target speaker.
When the length of the enrollment utterance is further increased from $4.0$ to $6.0$ seconds, the performance is not improved and appears saturated.
\PJHL{
The lack of monotonic improvement with longer enrollments is possibly due to the statistics of the enrollment lengths of the WSJ0-2mix dataset.
After SAD processing, the average enrollment length is about $4.53$ seconds in the training set and $5.57$ seconds in the test set, and therefore extending the enrollment length beyond their natural length often requires zero-padding, which adds no new speaker information.
Shorter segments might have already contain sufficient cues, while longer ones mainly incur padding and computation.
}

For TFLocoFormerV1, the performance is saturated when the enrollment utterance is $2$-second-long, and the model does not produce better performance when trained with longer enrollment utterances.

\begin{table}[t]
\scriptsize
\centering
\captionsetup{justification=centering}
\sisetup{table-format=2.2,round-mode=places,round-precision=2,table-number-alignment = center,detect-weight=true,detect-inline-weight=math}
\caption{\textsc{LExt Performance on WSJ0-2mix When Used with Various Enrollment Utterance Length.}}
\label{results_enrollment_length}
\setlength{\tabcolsep}{3pt}
\begin{tabular}{
c %
c %
c %
S [table-format=1.2,round-precision=2]
S [table-format=2.1,round-precision=1]
}
\toprule
{Row} & {System} & {DNN arch.} & {Prepend length (s)} & {SI-SDRi (dB)$\uparrow$} \\

\midrule

1Z & LExt & TFGridNetV1 & 0.1 & 18.717384699989733 \\
1A & LExt & TFGridNetV1 & 0.25 & 20.8392193402381 \\
1B & LExt & TFGridNetV1 & 0.5 & 21.97581145949584 \\
1C & LExt & TFGridNetV1 & 1 & 22.287854266446082 \\ 
1D & LExt & TFGridNetV1 & 2 & 22.770660203855176 \\
1E & LExt & TFGridNetV1 & 4 & \bfseries 23.015033027641476 \\
1F & LExt & TFGridNetV1 & 6 & 22.930041440859203 \\

\midrule

2A & LExt & TFLocoFormerV1 & 0.5 & 18.222826688789453 \\
2B & LExt & TFLocoFormerV1 & 1 & 20.150281953606367 \\
2C & LExt & TFLocoFormerV1 & 2 & \bfseries 20.260973487558463 \\
2D & LExt & TFLocoFormerV1 & 4 & 19.15245771238367 \\

\bottomrule
\end{tabular}

\vspace{0.3cm}

\scriptsize
\centering
\captionsetup{justification=centering}
\sisetup{table-format=2.2,round-mode=places,round-precision=2,table-number-alignment = center,detect-weight=true,detect-inline-weight=math}
\caption{\PJHL{\textsc{LExt Performance on WSJ0-2mix when Prepending vs. Both Prepending and Appending Enrollment Utterance.}}}%
\label{Prepending and Appending}
\setlength{\tabcolsep}{2.5pt}
\begin{tabular}{
c %
c %
S [table-format=1.1,round-precision=1]
S [table-format=1.1,round-precision=1]
S [table-format=2.1,round-precision=1]  %
S [table-format=2.1,round-precision=1]
}
\toprule
\multirow{2}{*}{System} & \multirow{2}{*}{DNN arch.} & 
{\multirow{2}{*}{\makecell[c]{Prepend\\length (s)}}} &
{\multirow{2}{*}{\makecell[c]{Append\\length (s)}}} & 
\multicolumn{2}{c}{SI-SDRi (dB)$\uparrow$} \\
\cline{5-6}
& & & & {Validation} & {Test} \\

\midrule

LExt & TFGridNetV1 & 4.0 & {-} & 22.22908498708671&\bfseries 23.015033027641476 \\ 
LExt & TFGridNetV1 & 2.0 & 2.0 & 22.082395720883273&22.90453079332287 \\
LExt & TFGridNetV1 & {-} & 4.0 & 22.206452915132047&22.77255337442582 \\
\midrule

LExt & TFLocoFormerV1 & 4.0 & {-} & 19.655&\bfseries19.22852909294151 \\
LExt & TFLocoFormerV1 & 2.0 & 2.0 & 18.960&18.003071901613634 \\
LExt & TFLocoFormerV1 & {-} & 4.0 &  19.225127987813902&17.771236281875947\\
LExt & TFLocoFormerV1 (NoPE) & 4.0 & {-} &  19.790534000702202&18.33268282750932\\
LExt & TFLocoFormerV1 (NoPE) & 2.0 & 2.0 & 19.704949682790787& 18.570565988621674\\
LExt & TFLocoFormerV1 (NoPE) & {-} & 4.0 & 19.92720839152171& 18.751277769266164\\

\bottomrule
\end{tabular}

\vspace{0.3cm}

\scriptsize
\centering
\captionsetup{justification=centering}
\sisetup{table-format=2.2,round-mode=places,round-precision=2,table-number-alignment = center,detect-weight=true,detect-inline-weight=math}
\caption{\textsc{LExt Performance on WSJ0-2mix when using V1 vs. V2 Models.}}
\label{results_simulated_delta}
\setlength{\tabcolsep}{2pt}
\begin{tabular}{
c %
c %
S [table-format=1.1,round-precision=1]
S[table-format=1.1,round-precision=1]
S[table-format=1.1,round-precision=1]
S[table-format=1.2,round-precision=2]
}
\toprule
{System} & {DNN arch.} & {Prepend length (s)} & {SI-SDRi (dB)$\uparrow$} & {SDRi (dB)$\uparrow$} & {PESQ$\uparrow$} \\

\midrule

Mixture & - & {-} & 0.0 & 0.0 & 1.6816486787597338 \\

\midrule

LExt & TFGridNetV1 & 4.0 & 23.015033027641476 & 23.16127470484226607 & 4.045368074536324 \\
LExt & TFGridNetV2 & 4.0 & \bfseries 24.100055543954173 & \bfseries 24.25952780624035607 & \bfseries 4.104941557864348 \\
[0.5ex]\hdashline\noalign{\vskip 0.5ex}
LExt & TFLocoFormerV1 & 4.0 & 19.22852909294151 & 19.50167961 & 3.7780266498128574 \\
LExt & TFLocoFormerV2 & 4.0 & \bfseries 21.921428360906916 & \bfseries 22.100151434197 & \bfseries 3.9763117094635962 \\

\bottomrule
\end{tabular}%
\end{table}

\subsection{\PJHL{Prepending vs. Appending vs. Both}}
In Table \ref{Prepending and Appending}, we compare the results of three strategies for concatenating enrollment utterance with mixture.
The first one prepends the enrollment utterance to the mixture, \PJHL{the second one appends the enrollment utterance to the mixture}, while the third one prepends the first half of the enrollment utterance while appends the second half.
In the third strategy, if the prepended segment is shorter than $E/2$, we pad zeros to its left, while if the segment concatenated on the right is shorter than $E/2$, we pad zeros to its right \PJHL{(i.e., the input structure is $[e_p; g; y; g; e_a]$, where $e_p$ and $e_a$ respectively denote the enrollment signals for prepending and appending).}
The rationale of doing so is to have the active speech in the enrollment utterance closer to the mixture.
\PJHL{Each strategy is applied consistently for training and testing.}

\PJHL{
In Table \ref{Prepending and Appending}, we do not observe better performance by using the second and third strategies, especially for TFLocoFormerV1.
This is possibly because, during training, a fixed-length segment is sampled from the original mixture and used for training, while, at run time, the input mixture is in its full length, which can be different from the segment length used for model training.
In this case, if the second and third strategies are used, the positional encodings of TF-LocoFormer used for the appended enrollment utterance would be different between training and testing, potentially creating difficulties for TF-LocoFormer to perform TSE. Therefore, we try a version without using positional encoding, denoted as ``NoPE'', in Table \ref{Prepending and Appending}. However, we observe that, despite using the NoPE version, TFLocoFormerV1 still shows no significant improvement.
We have also listed the TSE results on the validation set.
By comparing the metrics on the validation set and the test set, we find that TFGridNetV1 is more robust than TFLocoFormerV1, as the test scores of TFGridNetV1 in Table \ref{Prepending and Appending} are better than its validation scores, while the test scores of TFLocoFormerV1 are worse than its validation scores.

From Table \ref{Prepending and Appending}, we observe that the use of the first strategy performs better than the other two for both TFGridNetV1 and TFLocoFormerV1. Therefore, we adopt the first strategy in subsequent experiments.
It should be noted that we only compared their performance when the enrollment length is $4$ seconds, and their relative performance may change for other enrollment lengths.
}

\subsection{V1 vs. V2 Models of TF-GridNet and LocoFormer for LExt}

In Table \ref{results_simulated_delta}, based on the WSJ0-2mix dataset, we compare the TSE performance of using TFGridNetV1 and TFGridNetV2, and TFLocoFormerV1 and TFLocoFormerV2 in LExt.
We observe that the V2 models outperform V1, and TFGridNet outperforms TFLocoFormer.
We hence use TFGridNetV2 and prepend $4$-second enrollment utterance in default in the following experiments.

\subsection{LExt vs. Existing TSE Approaches}

Table \ref{results_compare_with_others} compares the result of LExt with the ones reported in previous studies, based on the WSJ0-2mix, WHAM!, and WHAMR! datasets.
We observe that LExt obtains clearly better TSE performance than existing approaches on all the three datasets, although being very simple.
Table \ref{results_compare_with_others} additionally lists speaker separation results (denoted as ``SS'') on the same datasets. We will discuss them later in Section \ref{TSE_vs_SS}.

WSJ0-2mix \cite{Hershey2016} is an anechoic dataset designed for two-speaker separation.
It does not contain environmental noises and room reverberation.
Notice that prepending an enrollment utterance to the mixture is equivalent to having the target speaker talk first and longer.
This onset information could help the DNN identify and extract the target speaker.
From the TSE results reported in Table \ref{results_compare_with_others}, we observe that LExt obtains state-of-the-art performance, reaching $24.1$ dB SI-SDR.
WHAMR! \cite{Maciejewski2020} is a noisy-reverberant dataset designed for two-speaker separation.
Since the enrollment utterance is anechoic without any noises and reverberation, the resulting concatenated mixture would exhibit some discrepancies between the signals in the time range of the enrollment utterance and those in the time range of the original mixture signal.
From the TSE results presented in Table \ref{results_compare_with_others}, we find that LExt achieves strong performance, obtaining $18.3$ dB SI-SDRi.
This indicates the effectiveness of LExt at TSE in noisy-reverberant conditions and that LExt can deal with the discrepancies.
We can draw similar conclusions based on the evaluation results on the WHAM! dataset \cite{Wichern2019}.
\PJHL{Although, in Table \ref{results_compare_with_others}, LExt only exploits $4$-second active enrollment speech rather than the full-length enrollment, it already shows better performance than the competing algorithms, which exploit full-length enrollment. We also tried to use full-length enrollment for LExt, but the model is too slow to train in the TFGridNetV2 setup and we hence leave it for future investigation.}

\begin{table*}[t]
\scriptsize
\centering
\captionsetup{justification=centering}
\sisetup{table-format=2.2,round-mode=places,round-precision=2,table-number-alignment = center,detect-weight=true,detect-inline-weight=math}
\caption{\textsc{SI-SDRi (dB)$\uparrow$, SDRi (dB)$\uparrow$ and PESQ$\uparrow$ Comparison of LExt with Existing Algorithms Based on WSJ0-2mix, WHAM! and WHAMR!.}}
\label{results_compare_with_others}
\setlength{\tabcolsep}{3pt}
\begin{tabular}{
l %
c %
c %
S[table-format=1.1,round-precision=1]
S[table-format=1.1,round-precision=1]
S[table-format=1.2,round-precision=2]
S[table-format=1.1,round-precision=1]
S[table-format=1.1,round-precision=1]
S[table-format=1.2,round-precision=2]
S[table-format=1.1,round-precision=1]
S[table-format=1.1,round-precision=1]
S[table-format=1.2,round-precision=2]
}
\toprule
 & & & \multicolumn{3}{c}{WSJ0-2mix} & \multicolumn{3}{c}{WHAM!} & \multicolumn{3}{c}{WHAMR!} \\
\cmidrule(lr){4-6} \cmidrule(lr){7-9} \cmidrule(lr){10-12}
{Row} & {System} & {Task} & {SI-SDRi$\uparrow$} & {SDRi$\uparrow$} &{PESQ$\uparrow$} & {SI-SDRi$\uparrow$} & {SDRi$\uparrow$} & {PESQ$\uparrow$} & {SI-SDRi$\uparrow$} & {SDRi$\uparrow$} & {PESQ$\uparrow$} \\

\midrule

0 & Mixture & - & 0.0 & 0.0 & 1.6816486787597338 & 0.0 & 0.0 & 1.4310215785106024 & 0.0 & 0.0 & 1.4152967761357624 \\

\midrule

1A & Conv-TasNet \cite{Luo2019} & \multirow{15}{*}{SS} & 15.3 & 15.6 &{-} & 12.7& {-} & {-} & 8.3& {-} & {-} \\
1B & DPRNN \cite{Luo2020} & & 18.8 & 19.0 & {-} & 13.9 & {-}&{-} & 10.3 &{-} & {-} \\
1C & WaveSplit \cite{Zeghidour2020} & & 21.0&21.2&{-} & 15.4 & 15.8 &{-} & 12.0 & 11.1 &{-} \\
1D & WaveSplit+DM \cite{Zeghidour2020} & & 22.2&22.3&{-} & 16.0&16.5&{-} & 13.2 & 12.2 & {-} \\
1E & SepFormer \cite{Subakan2021} & & 20.4 &20.5&{-}& 14.7 & 15.1&{-}&11.4 &10.4&{-} \\
1F & SepFormer+DM \cite{Subakan2021} & & 22.3 & 22.4&{-}&16.4&16.7 & {-}& 14.0 & 13.0 &{-} \\
1G & QDPN \cite{QDPN} & & 22.1 &{-}&{-}&{-}&{-}&{-} & 13.1&{-}&{-} \\
1H & DPTNet \cite{DPTnet} & & 20.2 &{20.6}&{-}&{-}&{-}&{-} & {-}&{-}&{-} \\
1I & MossFormer(L)+DM \cite{mossformer} & & 22.8&{-}&{-} & 17.3&{-}&{-} & 16.3&{-}&{-}\\
1J & MossFormer2(L)+DM \cite{mossformer2} & &  24.1 &{-}&{-}& 18.1 &{-}&{-} & 17.0 &{-}&{-}\\
1K & TD-Conformer-XL \cite{TD-conformer} & & 20.4 & {-}& {-} & {-} & {-} & {-} &13.1 &{-} &{-}\\ 
1L & TF-GridNet \cite{Wang2023TFGridNet} & & 23.5 & 23.6 & {-} & {-} & {-} & {-} &  17.1 & 15.6 & 2.69 \\
1M & TF-CrossNet \cite{Kalkhorani2024TFCrossNet} & & 23.2 & 23.4 & {-} & {-} & {-} & {-} & 17.9 & 16.4 & 2.91 \\ 

\midrule

2A & SpEx+ \cite{Ge2020SpEx+} & \multirow{12}{*}{TSE} & 16.9 & 17.2 & 3.43 & 13.1 & 13.6 &{-} & 10.9 & 10.0 & {-} \\
2B & DPRNN-Spe-IRA \cite{dprnn-ira} & & 17.3&17.6 & 3.43 & 14.15 & 14.61 & 2.57 & {-} & {-} & {-}\\ 
2C & SpEx++ \cite{spex++}& & 17.9 & 18.3 & 3.52 & 14.0 & 14.4 & {-} & 11.4 & 10.4 & {-}\\
2D & VEVEN \cite{Yang2023VEVEN}& & 19.0 & 19.2 & {-} & {-} & {-} & {-} & {-} & {-} & {-}\\
2E & X-TF-GridNet \cite{Hao2024XTFGridNet} & & 20.7 & 21.7 & 3.77 & 15.3 & 15.8 & {-} & 14.6 & 13.8 & {-} \\
2F & X-CrossNet\cite{x-crossnet} & & 19.9&20.5&{-} &{-}&{-}&{-}&14.6&14.1&{-}\\
2G & X-SepFormer \cite{x-sepformer} & & 18.9 &19.1&3.74& {-}&{-}&{-} & {-}&{-}&{-} \\
2H & CIENet-mDPTNet \cite{CIENet} & & 21.4&21.6&3.91 & 16.6& 17.0 & 2.70 & 15.7 &14.3&2.55 \\
2I & CIENet-C2F-mDPTNet \cite{CIENet-c2f} & &21.9 &22.1&3.94& 17.3&17.6&2.77 & 17.5&16.0&2.72 \\
2J & CIENet-Enh-mDPTNet \cite{CIENet-enh}& &21.5&21.8&{-} & 17.2 & 17.5 &{-}& 17.2&15.8&{-} \\
2K & DCF-Net \cite{DCFnet} & & 21.6&21.7&{-}& 16.8&17.3&{-}& 15.8 &14.5&{-}\\
2L & USEF-TSE \cite{Zeng2024USEF-TSE} & & 23.3&23.5&{-} & 17.6&17.9&{-} & 16.1&14.9&{-} \\

\midrule

3 & LExt-4s (TFGridNetV2) & TSE & \bfseries 24.100055543954173 & \bfseries 24.25952780624035607 & \bfseries 4.104941557864348 & \bfseries 18.329171820639434 & \bfseries 18.633428742455379 & \bfseries 2.9734255876342455 & \bfseries 18.292642083025615 & \bfseries 16.732450569684851 & \bfseries 2.938928936700026\\

\bottomrule
\multicolumn{12}{l}{\textit{Notes}: Systems trained with dynamic mixing are marked by ``DM''.}\\
\end{tabular}
\end{table*}

\begin{table}[t]
\scriptsize
\centering
\captionsetup{justification=centering}
\sisetup{table-format=2.2,round-mode=places,round-precision=2,table-number-alignment = center,detect-weight=true,detect-inline-weight=math}
\caption{\textsc{Performance Comparison of TSE for 2-speaker Mixtures with Same and Different Genders. Results are Based on WSJ0-2mix.}}
\label{results_gender}
\setlength{\tabcolsep}{6pt}
\begin{tabular}{
l
S[table-format=2.1,round-precision=1]
S[table-format=2.1,round-precision=1]
S[table-format=1.2,round-precision=2]
S[table-format=1.2,round-precision=2]
}
\toprule
 & \multicolumn{2}{c}{SDR (dB)$\uparrow$} & \multicolumn{2}{c}{PESQ$\uparrow$} \\
\cmidrule(lr){2-3}\cmidrule(lr){4-5}
{System} & {Different} & {Same} & {Different} & {Same} \\

\midrule

Mixture  & 2.50 & 2.70 & 2.29 & 2.34 \\

\midrule

SpeakerBeam \cite{Zmolikova2019SpeakerBeam} & 12.01 & 6.87 & 2.82 & 2.43  \\
SpEx \cite{Xu2020SpEx}  & 19.28 & 14.72 & 3.53 & 3.16  \\
SpEx+ \cite{Ge2020SpEx+} & 19.28 & 14.72 & 3.53 & 3.16  \\
SEF-Net \cite{Zeng2023SEF-Net}  & 20.20 & 18.73 & 3.56 & 3.48 \\
USEF-TSE \cite{Zeng2024USEF-TSE}  & 24.83 & 24.41 & 3.96 & 3.88 \\

\midrule

LExt (TFGridNetV2) & \bfseries 25.64041738467189 & \bfseries 25.09004811262652  & \bfseries 4.193595021572696 & \bfseries 4.163926940691326\\

\bottomrule
\multicolumn{5}{l}{\textit{Notes}: Following the setup in Table VII of the USEF-TSE paper \cite{Zeng2024USEF-TSE},}\\
\multicolumn{5}{l}{for each mixture in the test set, we only report the TSE result for the}\\
\multicolumn{5}{l}{first speaker, which always has higher energy (this is why the mixture}\\
\multicolumn{5}{l}{SDRs are much higher than $0$ dB).}
\end{tabular}%
\end{table}

In Table \ref{results_gender}, based on the two-speaker mixtures in the WSJ0-2mix test set, we compare the performance of LExt and existing studies for mixtures consisting of same- and different-gender speaker signals.
In both cases, we observe that LExt obtains better TSE performance than existing studies reporting the result in the same setup.

\begin{figure}
    \centering
    \includegraphics[width=8cm]{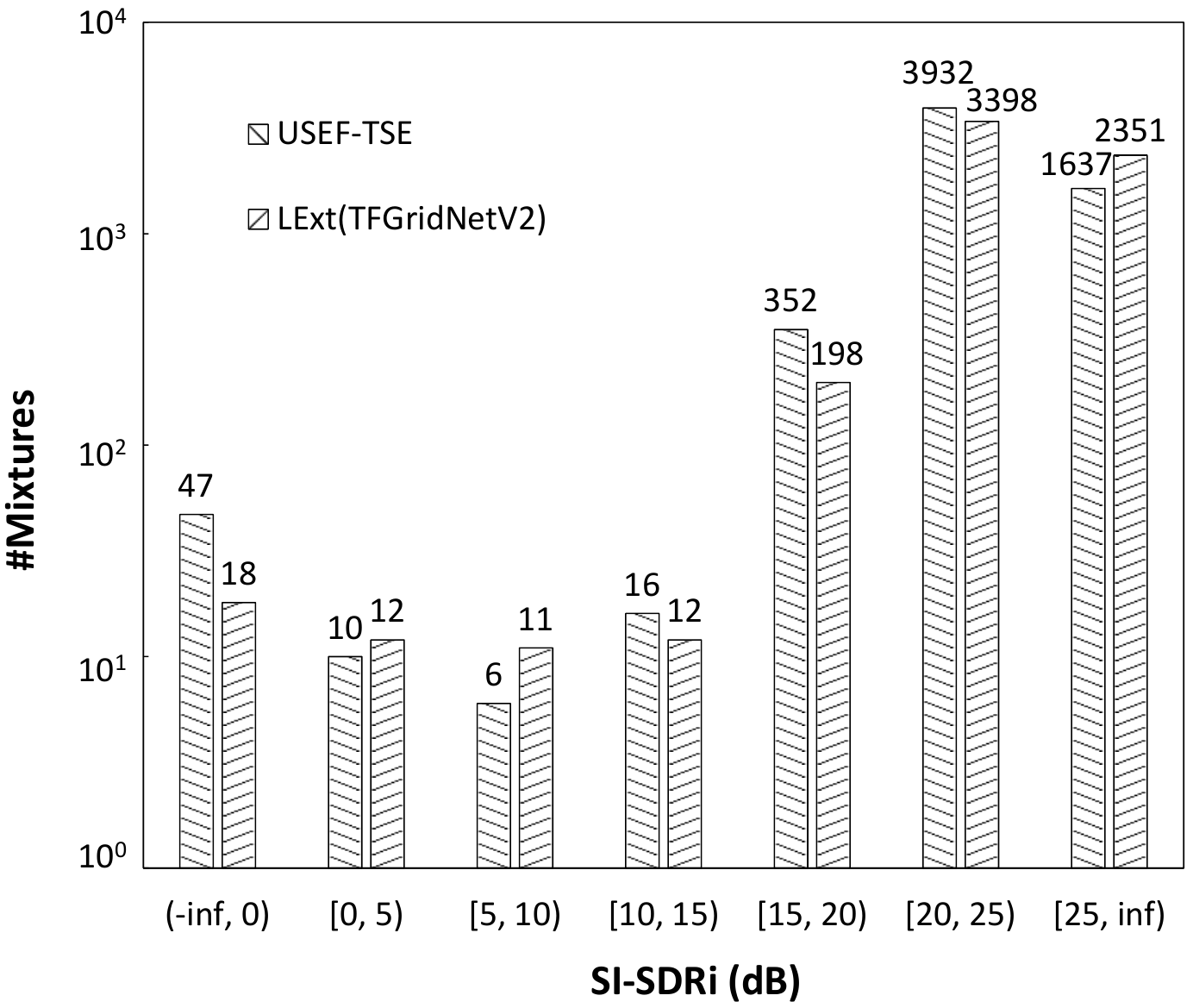}
    \caption{
    Histogram of SI-SDRi scores on WSJ0-2mix test set.
    ``Inf'' denotes infinity.}
    \label{distrbution_of_SISDRi_on_WSJ0-2mix}
\end{figure}

Fig. \ref{distrbution_of_SISDRi_on_WSJ0-2mix} plots the histograms of the SI-SDRi scores of USEF-TSE (in row 2L of Table \ref{results_compare_with_others}) and LExt (in row 3 of Table \ref{results_compare_with_others}) on the WSJ0-2mix test set.
We consider the cases when the SI-SDRi score is less than $0$ dB as failed cases (i.e., the left-most bar), and observe that LExt has fewer failed cases than USEF-TSE.
In addition, we observe that LExt has more mixtures with more than $25.0$ dB improvement in SI-SDR.

\subsection{Sensitivity to Length of Enrollment Utterance}

In Table \ref{sensitivity_enrollment_length}, we compare the performance of LExt, a TSE system based on fixed-length speaker embeddings, and USEF-TSE (based on variable-length speaker embeddings) when trained with various enrollment utterance lengths.
This comparison can show the sensitivity of various approaches to the length of enrollment utterances.
It can reveal which approach needs the least amount of enrollment speech.

\begin{table*}[t]
\scriptsize
\centering
\captionsetup{justification=centering}
\sisetup{table-format=2.2,round-mode=places,round-precision=2,table-number-alignment = center,detect-weight=true,detect-inline-weight=math}
\caption{\textsc{SI-SDRi (dB) Comparison of LExt vs. TSE with Fixed-Length Speaker Embeddings and USEF-TSE Based on WSJ0-2mix.}}
\label{sensitivity_enrollment_length}
\setlength{\tabcolsep}{5pt}
\begin{tabular}{
c %
c %
S [table-format=2.1,round-precision=1]
S [table-format=2.1,round-precision=1]
S [table-format=2.1,round-precision=1]
S [table-format=2.1,round-precision=1]
S [table-format=2.1,round-precision=1]
S [table-format=2.1,round-precision=1]
S [table-format=2.1,round-precision=1]
}
\toprule
& & \multicolumn{7}{c}{Length of enrollment utterance (s)} \\
\cmidrule(lr{9pt}){3-9}
 &  & \multicolumn{6}{c}{with SAD} & \multicolumn{1}{c}{without SAD} \\
\cmidrule(lr{9pt}){3-8}\cmidrule(lr{9pt}){9-9}
{System} & {DNN arch.} & 
0.1&{0.25} & 0.5 & 1.0 & 2.0 & 4.0 & {Full}  \\
\midrule

{TSE with fixed-length speaker embeddings \cite{Wang2024WeSep}} & {TFGridNetV1} &{Failed}&14.98834793188642 & 14.52668442605204 & 18.001471742712894 & 19.094972404710717 & 18.636276567353246 & 20.139508630490162 \\
{USEF-TSE (with variable-length speaker embeddings) \cite{Zeng2024USEF-TSE}} & {TFGridNetV1} &17.661015560495823& 19.48402791011865 & 20.37177037451033 & 21.819284293334157 & 22.19121682849402 & 22.91022920604795 & 22.67334340485977 \\

\midrule

{LExt} & {TFGridNetV1} &\bfseries 18.717384699989733 & \bfseries 20.8392193402381 & \bfseries 21.97581145949584 & \bfseries 22.287854266446082 & \bfseries 22.770660203855176 & \bfseries 23.015033027641476 & {\xmark} \\

\bottomrule
\end{tabular}
\vspace{-0.2cm}
\end{table*}

We leverage the implementation open-sourced in the WeSep toolkit \cite{Wang2024WeSep}
to build the TSE system based on fixed-length speaker embeddings.
See the details in Section \ref{baseline}.
For the USEF-TSE system, we leverage the USEF-TSE implementation\footnote{Available at \url{https://github.com/ZBang/USEF-TSE}.} released by the original authors \cite{Zeng2024USEF-TSE}.
The three systems in this experiment are all configured to use TFGridNetV1 as the speaker extraction network.
They are trained using exactly the same training configuration, and the only difference is in the way speaker conditioning is applied.

TSE systems based on fixed-length speaker embeddings and USEF-TSE, in default, both use each enrollment utterance in its full length for training and evaluation.
We denote this way as \textit{Full without SAD}.
In addition, after applying SAD to the enrollment utterances, for both training and testing we fix the enrollment length at $E$ seconds, where $E$ enumerates the set of $\{0.1, 0.25, 0.5, 1.0, 2.0, 4.0\}$, and report the results in Table \ref{sensitivity_enrollment_length}.

From the results, we observe that for various lengths of enrollment utterances, LExt obtains consistently better TSE performance than the other two.
In addition, LExt works reasonably well even if the enrollment utterance is as short as $0.5$ seconds, while in this case the TSE system based on fixed-length speaker embeddings performs significantly worse (i.e., $22.0$ vs. $14.5$ dB SI-SDRi).
\PJHL{
The degradation of the TSE model based on fixed-speaker embedding could be attributed to its pooling operation to get fixed-length speaker embedding: when the enrollment is very short, the aggregated embedding may become unstable and less reliable. 
For USEF-TSE, the enrollment and mixture are fused only once at the input of the network. As the network goes deeper, the enrollment cues may gradually fade away, and this effect is exacerbated when the enrollment itself is very short.
In contrast, LExt directly concatenates the enrollment with the mixture, so the speaker information is likely to remain salient throughout the layers. This design may be responsible for why LExt can maintain stronger TSE performance even in the extremely-short $0.5$-second enrollment condition.
}

\subsection{Illustration of Attention Maps}

\begin{figure}[t]
    \centering
    \includegraphics[width=8.5cm]{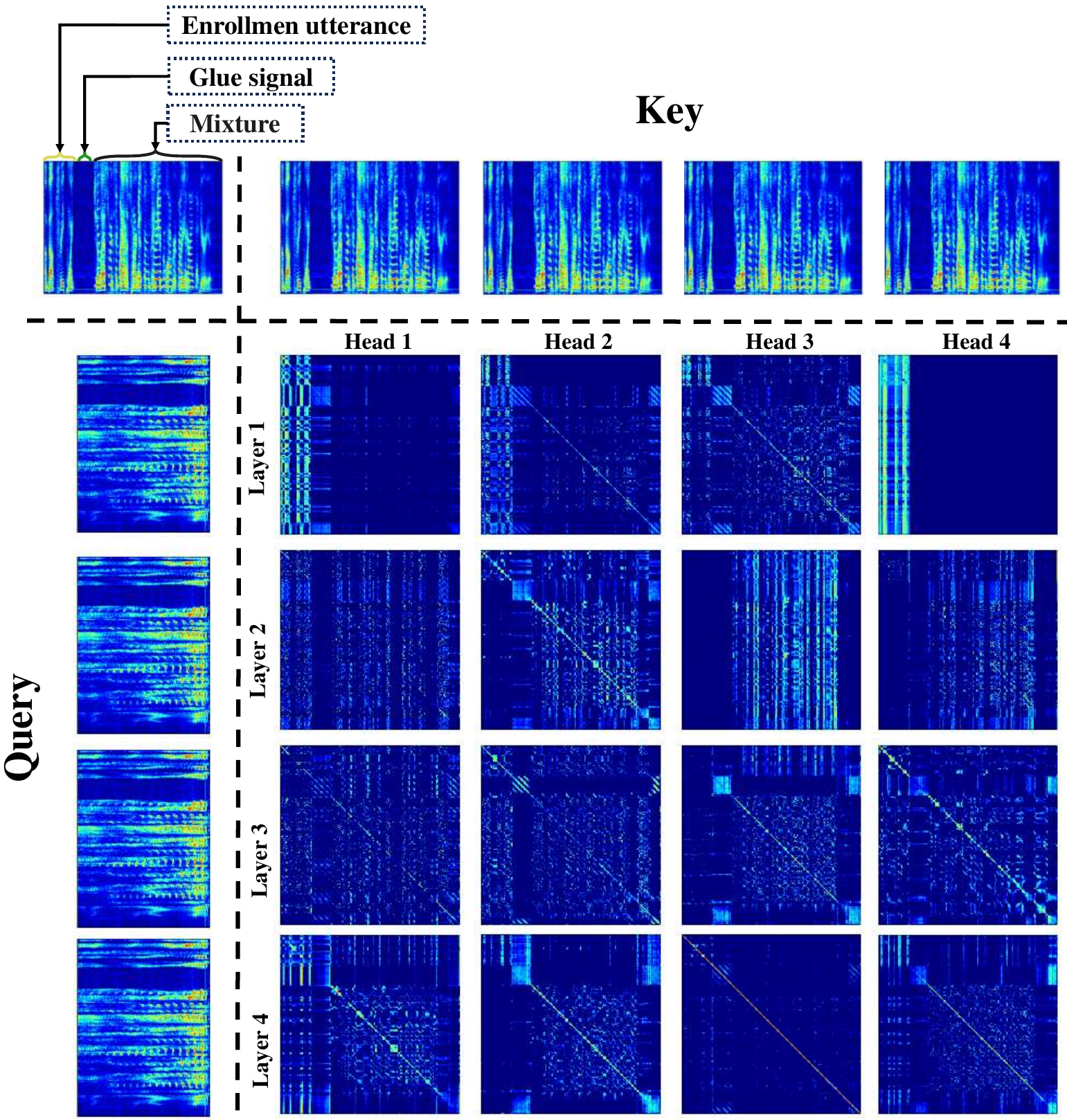}
    \caption{Illustration of attention maps in TFGridNetV1 based LExt reported in row 1C of Table \ref{results_enrollment_length}. In the top-left sub-plot, we mark down the time ranges of the enrollment utterance, glue signal and mixture. Best viewed in color.}
    \label{attn_map_illustration}
    \vspace{-0.2cm}
\end{figure}

Fig. \ref{attn_map_illustration} visualizes the attention maps of the TFGridNetV1 based LExt system reported in row 1C of Table \ref{results_enrollment_length}, based on a test mixture in WSJ0-2mix.
In TFGridNetV1, there are $4$ self-attention layers, each with $4$ attention heads, and Fig. \ref{attn_map_illustration} plots all the $16$ ($=4\times 4$) attention maps.
We observe that TFGridNetV1 indeed leverages self-attention to exploit the speaker cues in the time range of the prepended enrollment utterance, mainly in the first self-attention layer.

\subsection{TSE vs. Speaker Separation}\label{TSE_vs_SS}

In datasets designed for two-speaker separation (e.g., WSJ0-mix, WHAM! and WHAMR!), it is commonly observed in many TSE studies \cite{Elminshawi2022NewInsightsTSE} that first separating all the speakers and then, in an oracle way, aligning the speaker estimates to target speakers (denoted as \textbf{SS}, meaning ``speaker separation'') can yield better estimation of the target speakers than applying TSE to extract each of the target speakers in turn.
This can be observed in Table \ref{results_compare_with_others} by, e.g., (a) comparing row 2E and 2L with 1L, where X-TF-GridNet \cite{Hao2024XTFGridNet} and USEF-TSE \cite{Zeng2024USEF-TSE} designed for TSE are both not better than TF-GridNet \cite{Wang2023TFGridNet} designed for SS; (b) comparing row 1M and 2F, where X-CrossNet \cite{x-crossnet} designed for TSE is not better than TF-CrossNet \cite{Kalkhorani2024TFCrossNet} designed for SS; and (c) comparing row 1E with 2G, where X-SepFormer \cite{x-sepformer} designed for TSE is worse than SepFormer designed for SS \cite{Subakan2021}.
This indicates that many speaker conditioning mechanisms cannot sufficiently exploit speaker cues in enrollment utterances for TSE.
Comparing row 3 (TF-GridNet and LExt based TSE) with 1L (TF-GridNet based SS), we observe that LExt based TSE outperforms SS.
This comparison indicates that LExt is a very effective mechanism at exploiting speaker cues in enrollment utterances for TSE.

\PJHL{
\subsection{Effects of Utilizing Glue Signal}\label{glue_signal}

To assess the contribution of the glue signal, we conduct an ablation study by removing it from the LExt input. Recall that the glue signal is introduced to provide an explicit temporal cue that can indicate the boundary between the enrollment and mixture. Since the enrollment is fixed in length and in this case the boundary information is already clear, it is important to verify whether this explicit temporal cue can still play a useful role.
In addition, we investigate the impact of using different sample values for the glue signal.
Since the glue signal is only used to delimit the enrollment utterance from the mixture signal, we set its sample values to $5.0$.
The rationale of setting them to $5.0$ is because we perform gain normalization on both the enrollment and the mixture before feeding them to the network (see Section \ref{gain_norm}), which means that the sample values of the enrollment and mixture (after gain normalization) will hardly reach $5.0$, and therefore a signal with sample values set to $5.0$ can be a very informative delimiter.

The results are shown in Table \ref{tab:glue_value_length}.
Firstly, we observe that setting glue values to $5.0$ can produce better results, indicating that setting glue values to $5.0$ can better serve as a delimiter between enrollment and mixture.
Secondly, we find that the glue signal does not play a decisive role in the network's differentiation between enrollment and mixture. When we use a fixed-length enrollment, the network can still determine through training that the fixed-length signal at the front is the enrollment. However, when the length of the enrollment is not fixed and the original, full-length enrollment utterance is used, the presence of the glue signal enables the network to better distinguish between enrollment and mixture.
}

\begin{table}[t]
\scriptsize
\centering
\captionsetup{justification=centering}
\sisetup{
  table-number-alignment = center,
  round-mode = places,
  round-precision = 1,
  detect-weight = true,
  detect-inline-weight = math
}
\caption{\PJHL{\textsc{Ablation Results of LExt with and without using Glue Signal, and Using Different Glue Values.
Experiments are Based on WSJ0-2mix and TFGridNetV1.}}}
\label{tab:glue_value_length}
\begin{tabular}{
  c %
  c %
  c %
 S[table-format=2] %
  S[table-format=2.1] %
}
\toprule
{ID} & {Enrollment} & {Glue Value} & {Glue Length (ms)} & {SI-SDRi (dB)$\uparrow$} \\
\midrule
1 & Fixed at 1s & {-} & 0  & 22.40154945945212 \\
2 & Fixed at 1s & $0.0$ & 32 & 22.287854266446082 \\
3 & Fixed at 1s & $5.0$ & 32 & \bfseries 22.54095719500445 \\
\midrule
4 & Full & {-} & 0  & 20.479600577846796 \\
5 & Full & $0.0$ & 32 & 22.614180557333555 \\
6 & Full & $5.0$ & 32 & \bfseries 23.02442964177268 \\
\bottomrule
\end{tabular}
\end{table}

\PJHL{
\subsection{Effects of Applying SAD to Enrollment}\label{SAD}
We study how SAD would affect LExt across different enrollment lengths, including very short lengths ($0.25$\,s and $0.5$\,s) and longer ones ($1$\,s and $2$\,s).
For each length, we compare models trained and evaluated without applying SAD and with applying SAD to the enrollment.

From the results in Table \ref{tab:sad_vs_len}, we observe that for short enrollments, when SAD is not used, there is a significant drop in TSE performance. Specifically, when the enrollment length is as short as $0.25$ s, the performance of the ``without SAD'' setting drops drastically: its SI-SDRi value is only at $9.7$ dB, while the ``with SAD'' setting achieves an SI-SDRi of $20.8$ dB.
This is because during testing, when we cut a short segment from the entire enrollment speech for LExt, without SAD, we could obtain a silent segment or a segment without too much active speech, which would lead to extraction failures.
However, as the signal length increases, the performance with and without SAD becomes relatively close.
}

\begin{table}[t]
\scriptsize
\centering
\captionsetup{justification=centering}
\sisetup{
  table-number-alignment = center,
  round-mode = places,
  detect-weight = true,
  detect-inline-weight = math
}
\caption{\PJHL{\textsc{Effects of SAD at Different Enrollment Lengths. Experiments are based on WSJ0-2mix and TFGridNetV1.} 
$\Delta$ = (SI-SDRi with SAD) $-$ (SI-SDRi w/o SAD).}}
\label{tab:sad_vs_len}
\begin{tabular}{
  S[table-format=1.2,round-precision = 2]  %
  S[table-format=2.1,round-precision = 1]  %
  S[table-format=2.1,round-precision = 1]  %
  S[table-format=+1.1,round-precision = 1] %
}
\toprule
{Length\,(s)} & {SI-SDRi (dB) w/o SAD} & {SI-SDRi (dB) with SAD} & {$\Delta$} \\
\midrule
0.25 &  9.72852326953706 & 20.8392193402381 & 11.11069607070104 \\
0.50 &  19.808626379287258 & 21.97581145949584 & 2.167185080208582 \\
1.00 &  22.56084025426923 & 22.287854266446082 & -0.272985987823148 \\
2.00 &  22.54913818482465 & \bfseries 22.770660203855176 & 0.221522019030526 \\
\bottomrule
\end{tabular}
\end{table}

\section{\PJHL{Disussions}}\label{limitations}

This section describes several limitations of LExt compared with fixed-length speaker embedding based systems for TSE.

\subsection{Limitations of LExt for Offline TSE}\label{limitations_offline}

LExt increases the input length by concatenating the enrollment utterance with the mixture, which leads to higher computational cost during TSE.
The amount of increased computation is in proportion to the length of the concatenated enrollment utterance.
However, this may not be a serious problem, as in our experiments we find that an enrollment utterance as short as $0.25$ or $0.5$ seconds (see Table \ref{results_enrollment_length}) can already produce very strong extraction performance.
This length is much shorter than that of typical mixture signals, which is typically multi-seconds long.

Another related limitation is that every time LExt is used for inference, the enrollment utterance has to be processed together with the mixture signal.
We cannot store the processing results of the enrollment utterance on hard drives for later use, as they are different for different mixture signals.
This is a weakness compared to TSE systems based on fixed-length speaker embeddings, where a speaker embedding is extracted only once for each speaker and can be stored on disks, or in cache, for later use.

\subsection{Limitations of LExt for Real-Time, Frame-Online TSE}\label{limitations_online}

Another limitation may emerge when applying LExt for real-time, frame-online TSE, since, during inference, the DNN needs to refer to the enrollment utterance concatenated at the very beginning of the mixture for TSE.
In real-time scenarios, where the computation resource is often limited, referring to the tensors computed based on the enrollment utterance by using, e.g., attention mechanisms would be costly.
Future research will explore solutions to this problem. 

On the other hand, in online TSE, we can cache the tensors computed from the enrollment utterance (assuming only using prepending in LExt), and re-use them to extract the target speaker in different mixtures.
That is, the second limitation described in Section \ref{limitations_offline} can be avoided in online TSE.

\section{Conclusions}\label{conclusion}

We have proposed LExt for TSE.
By concatenating an enrollment utterance to the mixture to create an artificial onset for the target speaker, LExt is found highly-effective at extracting the target speaker, achieving state-of-the-art TSE performance on multiple widely-adopted datasets.
Even if the enrollment utterance used for concatenation is as short as $0.25$ or $0.5$ seconds, LExt can still obtain reasonably strong TSE performance.
LExt is extremely simple to implement and does not require additional design of DNN architectures for speaker conditioning.
Moving forward, we plan to extend LExt for related target-speaker speech processing tasks and address its limitations described in Section \ref{limitations}.

In closing, we highlight that, in the past decade, TSE has been conducted mainly following the convention of designing additional DNN modules for speaker conditioning (e.g., by leveraging fixed- or variable-length speaker embeddings) so that the speaker extraction network can learn to extract the target speaker.
LExt, for the first time, proposes to use no additional speaker-conditioning modules, and the speaker extraction network can learn to leverage onset cues for TSE.
This approach, we think,
could potentially motivate the design of many TSE algorithms in future research.

\bibliographystyle{IEEEtran}
\bibliography{references.bib}

\end{document}